\documentclass{moriond}

\usepackage{amsmath}
\usepackage{amssymb}
\usepackage{amsfonts}
\usepackage{graphicx,bbm,mathrsfs}

\def\be{\begin{equation}}
\def\ee{\end{equation}}
\def\bea{\begin{eqnarray}}
\def\eea{\end{eqnarray}}

\newcommand{\Photo}{\includegraphics[angle=0,height=35mm]{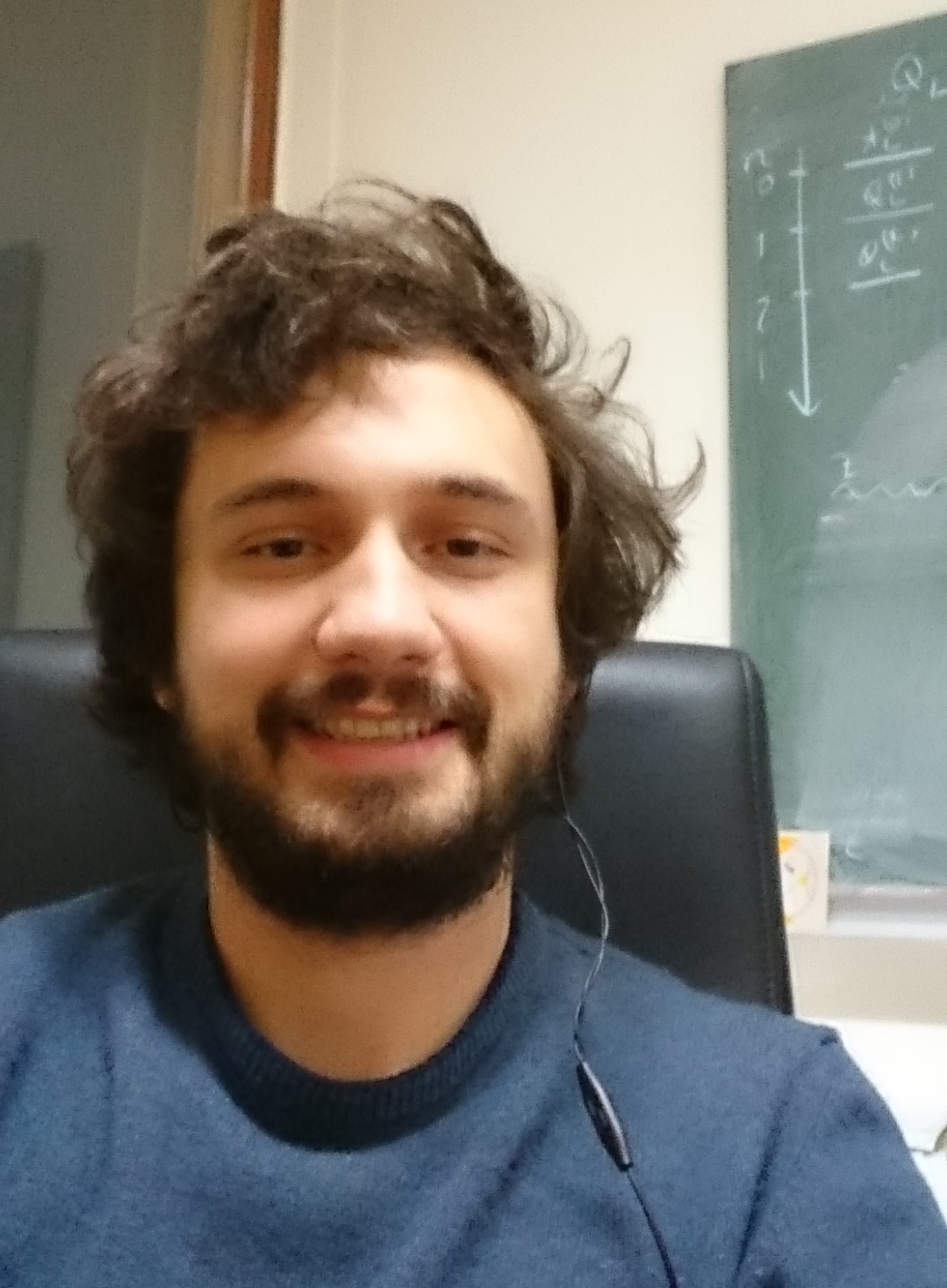}}

\begin{document}
\vspace*{4cm}
\title{SCALAR PRODUCTION IN ASSOCIATION WITH A Z BOSON AT LHC AND ILC: THE MIXED BEH-RADION CASE OF WARPED MODELS}

\author{ A. ANGELESCU }

\address{Laboratoire de Physique Th\'{e}orique, B\^at. 210, CNRS,
Univ. Paris Sud, \\  Universit\'{e} Paris-Saclay, F-91405 Orsay Cedex, France}

\maketitle\abstracts{
The radion scalar field might be the lightest new particle predicted by extra-dimensional extensions of the Standard Model, thus possibly leading to the first signatures of new physics at the LHC. We perform a study of its production in association with a $Z$ boson in the custodially protected warped model with a brane-localised Higgs boson addressing the gauge hierarchy problem, with radion-Higgs mixing effects included. While the considered radion production at the LHC would constrain some parts of the parameter space, it is only the ILC program that will be able to cover a significant part of this space through the studied process. Complementary tests of the same theoretical parameters can be realised through the high accuracy measurements of the Higgs couplings at ILC.}

\section{Introduction} \label{se:intro}

An elegant class of extensions of the Standard Model (SM) addressing the gauge hierarchy problem are the so-called warped extra-dimensional or Randall-Sundrum (RS) models.~\cite{RS} In such models, the hierarchy between the Planck and electroweak scales is achieved with the help of the curved geometry of the 5th dimension.

However, as the extra dimension is finite in length, it should be stabilized by a suitable mechanism. One such proposal is the Goldberger-Wise mechanism,~\cite{GW} which predicts the existence of a new scalar field, the radion, that can mix with the SM scalar boson. In the following, after summarizing the main properties of the radion, we will discuss the prospects of its detection when produced in association with a $Z$ boson at present (LHC) and future colliders (ILC).

\section{Higgs-Radion Mixing Couplings to Z Bosons} \label{sec:couplings}

The model under consideration is the custodial RS scenario~\cite{RScusto} with the Higgs bi-doublet localised on the infrared (IR) brane, while the remaining fermionic and gauge fields are propagating in the bulk. In the $(+----)$ convention that will be used throughout this work, the perturbed RS metric reads

\begin{equation}
{\rm d} s^2 = {\rm e}^{-2(k\,y+F)} \eta_{\mu\nu} {\rm d}x^{\mu}{\rm d}x^{\nu} - (1+2F)^2 {\rm d}y^2 \equiv  g_{MN} {\rm d}x^{M}{\rm d}x^{N},
\label{perturbed_metric}
\end{equation}
where $F(x,y)$ represents the scalar perturbation of the 5D metric, which we denote as $g_{MN}$.\footnote{Upper case roman letters denote 5D Lorentz indices, while the Greek letters denote 4D Lorentz indices.} Upon resolution of the 5D Einstein equations and in the limit of small backreaction of the field $F$ on the metric curvature, the scalar perturbation $F(x,y)$ can be parametrized as follows~\cite{RadHatIR}:
\begin{equation}
F(x,y) = \frac{\phi_0 (x)}{\Lambda} {\rm e}^{2 k (y-L)},
\label{radion_profile}
\end{equation} 
where $\phi_0$ is the (unmixed) 4D radion field and $\Lambda$ its vacuum expectation value (VEV), which is an $\mathcal{O}$~(TeV) energy scale that sets the length of the extra dimension.\cite{GW}

In order to obtain the (unmixed) radion coupling to two $Z$ bosons, one should linearize the corresponding 5D Lagrangian with respect to the scalar perturbation, $F$. This 
procedure, including the effects of Kaluza-Klein excitations
of the $Z$ boson, is performed in detail in reference~\cite{Angelescu:2017jyj}. Here we only display the dominant contributions to the $h_0 ZZ$ and $\phi_0 ZZ$ couplings, in the limit of infinitely heavy KK resonances:
\begin{equation}
\mathcal{L}_{\varphi Z Z}^{\rm 4D} \simeq m_Z^2 \left( \frac{h_0}{v} - \frac{\phi_0}{\Lambda
} \right) Z_{\mu} Z^{\mu},
\label{L_ZZ_scalar}
\end{equation} 
where $h_0$ denotes the unmixed Higgs scalar and $v$ its VEV.

The Higgs-radion mixing arises at the renormalisable level by coupling the 4D Ricci scalar $R_4$ to the trace of $H^{\dagger} H$ via a gauge invariant term~\cite{RadSMatIR} as follows:
\begin{equation}
S_{\xi}^{\rm 4D} = \xi \int {\rm d}^4 x \, \sqrt{g^{\rm ind}} \, R_{4 }(g^{\rm ind}_{\mu\nu}) \frac{1}{2} \, \mathrm{tr} \left( H^{\dagger} H  \right),  
\label{hr_mixing}
\end{equation}
with $g^{\rm ind}_{\mu\nu}$ being the perturbed metric induced on the IR brane metric. As it involves the brane-localised Higgs field, the Higgs-radion mixing sources from the IR brane. After EW symmetry breaking, a non-zero $\xi$ coupling in induces a kinetic mixing between $h_0$ and $\phi_0$. The transition to the mass eigenstates, $h$ and $\phi$, is worked out in detail in reference~\cite{Angelescu:2017jyj}. Here, we only mention that the Higgs-radion sector is described by four parameters: the mixing parameter $\xi$, the radion VEV $\Lambda$, the physical radion mass $m_{\phi}$, and the physical Higgs mass $m_h=125$~GeV.

\begin{figure}[t!]
\begin{center}
\hspace*{\fill}
\includegraphics[keepaspectratio=true,width=0.4\textwidth]{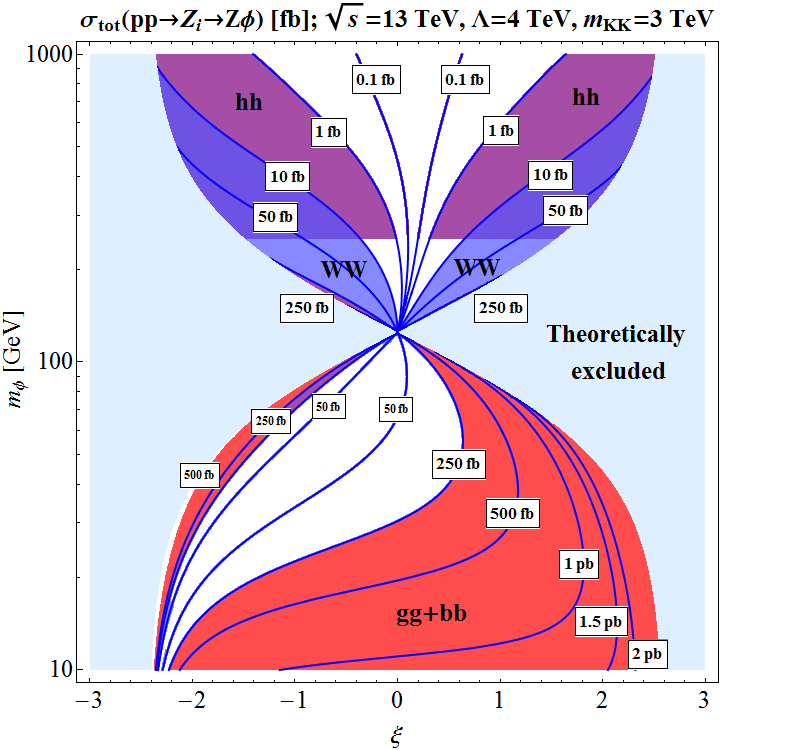}
\hfill
\includegraphics[keepaspectratio=true,width=0.37\textwidth]{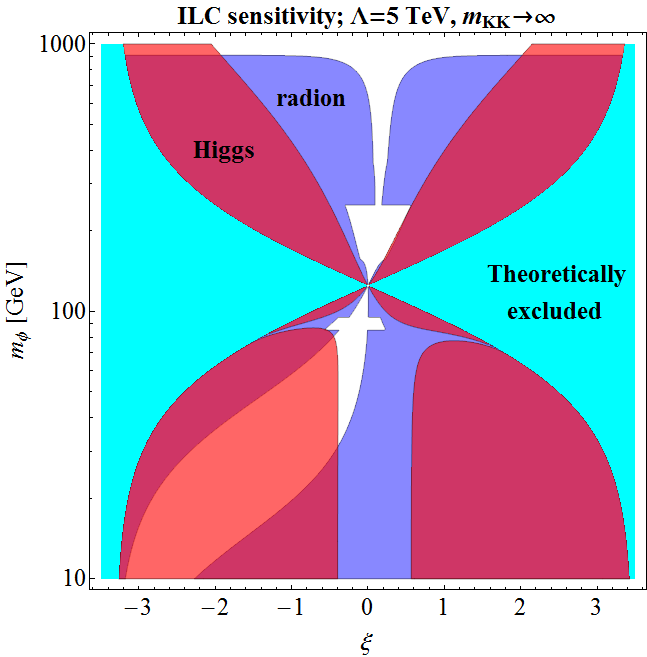}
\hspace*{\fill}
\caption{
{\small \underline{Left panel}: Iso-contours of $Z\phi$ production cross section (in fb and pb) at the LHC with $\sqrt{s} = 13$~TeV, as a function of $\xi$ and $m_{\phi}$ (in GeV), for $\Lambda=4$~TeV. The light blue regions are excluded by theoretical constraints, while the purple, red, and blue zones approximately indicate parameter space regions that will be probed with $300$~fb$^{-1}$ at the LHC via radion decays into $hh$, dijets ($gg+bb$), and $WW$ final states, respectively. \underline{Right panel}: Summary plot for direct and indirect radion searches at the three stages of operation of the ILC ($\sqrt{s} = 250$~GeV, $500$~GeV, and $1$~TeV), in the $\{\xi,m_{\phi}\}$ plane, for $\Lambda=5$~TeV. The blue region covers the Higgs-radion parameter space estimated to be probed by the ILC through direct radion searches, while the red region represents the domain potentially probed by the precise measurement of the $hZZ$ coupling. The theoretically excluded region is represented as the cyan domain.}
}
\label{fig:lhc_ilc}
\end{center}
\end{figure}

\section{Radion Detection and Higgs Precision Measurements}\label{sec:EXP}

In this section, we will summarize the search strategies for a radion produced in association with a $Z$ boson at the LHC and at the ILC, and then discuss how Higgs precision measurements at the ILC can constrain the mixed Higgs-radion parameter space. A more detailed discussion can be found in reference \cite{Angelescu:2017jyj}.

\subsection{Radion Production at the LHC} \label{sec:LHCapplied}

For the full reaction $pp \to Z\phi$ followed by the radion decay into a pair of SM states, one expects a considerable SM background that needs to be rejected. Such a reduction of the background can come from a cut on the transverse momentum of the reconstructed $Z$, $p_T(\mu\mu)$ (see the $p_T(\mu\mu)$ distribution for the 13 TeV LHC in reference~\cite{DY2jets}). We now summarize the main techniques to measure $Z\phi$ production at the LHC, focusing on three ranges for the radion mass.

\noindent\textbf{$\boldsymbol{m_{\phi} \gtrsim 20}$~GeV}. A light radion decays almost always into jets ($gg$ or $bb$ pairs), which corresponds to selecting experimentally two inclusive jets (including two gluons or two $b$'s). By appropriately choosing the cuts on the $p_T$ of the reconstructed $Z$,~\cite{Angelescu:2017jyj} the obtained LHC reach at $2\sigma$ for $300$~fb$^{-1}$ at 13 TeV is illustrated by the red region in the left panel of Fig.~\ref{fig:lhc_ilc}.

\noindent\textbf{$\boldsymbol{m_{\phi} > 160}$~GeV}. In this regime, one benefits from the kinematical opening of the $WW$ channel (the branching ratio to $ZZ$ is smaller). For an integrated luminosity of $300$~fb$^{-1}$, selecting semi-leptonic decays for the $WW$ system and corroborating with a radion mass selection, one obtains the sensitivity order of magnitude indicated by the blue region on the left panel of Fig.~\ref{fig:lhc_ilc}.

\noindent\textbf{$\boldsymbol{m_{\phi} > 250}$~GeV}. Finally, for $m_{\phi} > 250$~GeV, the LHC becomes sensitive to the channel $pp \to Z\phi$, $\phi \to hh$. The $Zhh$ production background opens up with a cross section of $0.25$~fb,~\cite{Zhhrate} which would correspond to a $\sim 1$~fb cross section sensitivity limit for $\sigma_{tot}(Z\phi)$. The order of magnitude of this sensitivity is indicated, in purple, on the left panel of Fig.~\ref{fig:lhc_ilc} as well.

This domain and the above sensitivity regions are clearly coarse estimates and a full analysis would be needed. Those regions however show that the $Z\phi$ search at LHC could be complementary, in testing some specific regions of the $\{\xi,m_{\phi}\}$ plane, to the search for the gluon-gluon fusion production of the radion.~\cite{Frank}

\subsection{Radion Production at the ILC} \label{sec:ILCapplied}

For the associated $Z\phi$ production at ILC, one can use the same missing mass technique as for the $Zh$ production~\cite{MissMass} which is independent of the radion branching ratio values. This powerful method is only feasible using the large luminosity provided by this machine (H-20 scenario~\cite{H20scen}). We summarize below the prospects for $Z+\phi$ production at the ILC for three representative mass ranges for the radion.

\noindent\textbf{$\boldsymbol{m_{\phi} < 150}$~GeV}. Since the $Z\phi$ production cross section scales as $s^{-1}$, this low mass domain will be covered by running at $\sqrt{s}=250$~GeV. The $\sigma(Z\phi)$ sensitivity in this mass region which is at the $0.5-1$~fb level, as shown by the blue region from the right panel of Fig.~\ref{fig:lhc_ilc}. Note the decrease of sensitivity for $m_{\phi} \sim m_Z$, where a radion peak would be obscured by the $Z$ peak.

\noindent\textbf{$\boldsymbol{m_{\phi} > 150-160}$~GeV}. When $m_{\phi} >150$~GeV, one starts crossing the kinematical limit for the $Z\phi$ production and it becomes necessary to use data taken at a $500$~GeV center-of-mass energy. Also, the situation changes substantially since the $WW$, $ZZ$ channels start opening for the radion decay, which helps the recoil techniques. Selecting semileptonic decays of the $WW$ system, one reaches, at the counting level, a $\sim 1$~fb sensitivity on $\sigma(Z\phi)$.

\noindent\textbf{$\boldsymbol{m_{\phi} > 250}$~GeV}. For $m_{\phi} >250$~GeV, the $hh$ channel becomes accessible for the radion decay. Using Higgs decays into $b\bar b$, one could reach a sensitivity on $\sigma(Z\phi)$ at the $\sim 0.01$~fb level.

The various estimates given so far constitute a reasonable first guess of the ILC sensitivity for a radion search in the $Z\phi$ production channel. On the right panel of Fig.~\ref{fig:lhc_ilc}, we summarize on a unique plot the covered regions (in blue) issued from the ILC runs at $250$~GeV, $500$~GeV and $1$~TeV, for the radion VEV having a value of $\Lambda=5$~TeV. A dedicated analysis would be needed to fully assess such performances but it is clear that ILC can dig into the radion scenario with excellent sensitivity.

\subsection{Higgs Precision Measurements at the ILC} \label{sec:HiggsProdILC}

It is estimated that the Higgs coupling to two $Z$ bosons will be measured at the $0.51\%$ ($1.3\%$) $1\sigma$ error level at the ILC with an energy option of $1$~TeV ($250$~GeV),~\cite{ILCperH} via the $Zh$ production channel. Such measurements would exclude at $2\sigma$ the red regions from the right panel of Fig.~\ref{fig:lhc_ilc}, assuming a central value equal to the predicted SM $hZZ$ coupling.

\section{Conclusions} \label{se:conclu}

Let us finish this study on the radion production by a short conclusion, now that the possibilities of observation have been discussed. The study of the reaction $q \bar q \to Z\phi$ at the LHC could allow to cover significant parts of the $h-\phi$ parameter space. However, it will take the ILC program at high luminosity to cover most of the theoretically allowed parameter space, via the $e^+e^- \to Z\phi$ search. Moreover, the ILC benefits from the complementarity between i) direct radion searches and ii) high accuracy measurements of the Higgs couplings in the exploration of the $h-\phi$ parameter space (typically the $\{\xi,m_{\phi}\}$ plane). We also note that, due to the better sensitivity of the ILC with respect to the LHC for the low radion masses, a light radion is an interesting example of new physics that could escape detection at the LHC but be discovered at the ILC.

\section*{Acknowledgments}

This work was supported by the ERC advanced grant ``Higgs@LHC''.

\end{document}